\makeatletter
\let\blx@rerun@biber\relax
\makeatother

\documentclass[conference,flushend]{iaria}

\usepackage{tikz}
\usetikzlibrary{arrows,positioning}
\usepackage{cleveref}
\usepackage{url}
\usepackage{hyperref}
\usepackage[hyphenbreaks]{breakurl}

\title{Graph of Effort: Quantifying Risk of AI Usage for Vulnerability Assessment}
\author{
  \IEEEauthorblockN{%
    Anket Mehra, Andreas Aßmuth\,\orcidlink{0009-0002-2081-2455}, and Malte Prieß\,\orcidlink{0009-0004-4626-2513}}
  \IEEEauthorblockA{%
    Department of Computer Science and Electrical Engineering\\
    Kiel University of Applied Sciences\\
    Kiel, Germany\\
    e-mail: {\tt anket.mehra@student.fh-kiel.de, $\lbrace$andreas.assmuth\,|\,malte.priess$\rbrace$@fh-kiel.de}
} }

\usepackage{ifpdf}
\ifpdf
\pdfoutput=1 % we are running pdflatex 
\pdftrue
\fi

\makeatletter
\def\ps@IEEEtitlepagestyle{
	\def\@oddfoot{\mycopyrightnotice}
	\def\@evenfoot{}
}
\def\mycopyrightnotice{
	{\footnotesize
		\begin{minipage}{0.8\textwidth}
			\centering
			Please cite as: Anket Mehra, Andreas Aßmuth, and Malte Prieß, ``Graph of Effort: Quantifying Risk of AI Usage For Vulnerability Assessment,'' in \emph{Proc of the 16th International Conference on Cloud Computing, GRIDs, and Virtualization (Cloud Computing 2025), p.~17--24, Valencia, Spain, April 2025.}
		\end{minipage}
	}
}
\makeatother
\begin{document}
\maketitle
\begin{abstract}
With AI-based software becoming widely available, the risk of exploiting its capabilities, such as high automation and complex pattern recognition, could significantly increase. An AI used offensively to attack non-AI assets is referred to as offensive AI. Current research explores how offensive AI can be utilized and how its usage can be classified. Additionally, methods for threat modeling are being developed for AI-based assets within organizations. However, there are gaps that need to be addressed. Firstly, there is a need to quantify the factors contributing to the AI threat. Secondly, there is a requirement to create threat models that analyze the risk of being attacked by AI for vulnerability assessment across all assets of an organization. This is particularly crucial and challenging in cloud environments, where sophisticated infrastructure and access control landscapes are prevalent. The ability to quantify and further analyze the threat posed by offensive AI enables analysts to rank vulnerabilities and prioritize the implementation of proactive countermeasures. To address these gaps, this paper introduces the Graph of Effort, an intuitive, flexible, and effective threat modeling method for analyzing the effort required to use offensive AI for vulnerability exploitation by an adversary. While the threat model is functional and provides valuable support, its design choices need further empirical validation in future work.
\end{abstract}
\begin{IEEEkeywords}
threat modeling; vulnerability assessment; offensive~AI.
\end{IEEEkeywords}

\section{Introduction}

At the latest since the presentation of ChatGPT (GPT-3.5) by OpenAI in November 2022, the topic of Artificial Intelligence (AI) has also been omnipresent in the general public. As we are now seeing, AI also has an impact on cybersecurity, as the availability of AI services can make known attacks easier or more efficient. The attack itself does not even have to be technical. An employee of a multinational company was tricked into making a bank transfer of $25.6$~million US dollars using AI-generated deepfakes \cite{ChenMagramo2024}. He had taken part in a video conference, supposedly together with the Chief Financial Officer and other employees of the company. In reality, however, all the other people were deepfakes. And this is not an isolated case: Hong Kong police announced that there have been other similar incidents in which deepfakes were used to deceive facial recognition programs.
The question arises as to how the availability of AI services and their use in cyberattacks can be quantified and how this aspect can be taken into account in known threat modeling methods.\par 

Several publications deal with the creation of AI models for cybersecurity. A comprehensive overview can be found in \cite{schroer2024sok}. Rising capabilities of AI lead to more complex models, advanced train infrastructure and accessible datasets. Importance for research and understanding it continuously is thus increasing. Since AI is a ``dual-use technology'' \cite{enisa2023}, its capabilities can be used in good or malicious directions. For example, using AI can help organizations to enhance their cyber defense mechanisms, but on the other hand and as indicated by the above given example, adversaries can use AI to simplify or improve attacks, as well as to create new and sophisticated attack vectors.

A current gap in research is the quantification of AI threats. To the best of our knowledge, only \cite{Mirsky_2023} tried to quantify the threat of offensive AI by a survey with industry experts. Furthermore, in \cite{Malatji_2024} it was identified that there are no methods to quantify factors for the AI threat, namely the motivation. Also, the applicability of AI security research is criticized by \cite{grosse2023machine} because most experiments rely on artificial environments, making the application of research results insufficient.  

The latest version 4.0 of the Common Vulnerability Scoring System (CVSS) contains the criterion ``automatable'' in the new (optional) Supplemental Metric Group. As the term indicates, this criterion is intended to assess the automation potential of an attack, however, it is not intended to address specifically AI-based automation \cite{First-cvss-4}. Furthermore, the CVSS base score examines the severity of a vulnerability by choosing scores between $0$ and $10$ for different a-priori categories, which are directly related to the vulnerability. External factors are included in the environmental and supplemental metrics. These can be used to quantify the risk of a vulnerability according to different IT system environments. However, in conclusion, none of the metrics examine the AI risk \cite{First-cvss-4}. 

Being able to quantify the AI threat is important. It allows cyber analysts to prioritize threats and provides support to explain the specific danger of them. Additionally, it allows organizations to proactively implement target-oriented countermeasures. 

Furthermore, the importance raises in cloud-based IT systems. These are characterized by having sophisticated connections between multiple components such as deployed web applications or the necessary infrastructure, such as databases and authorization management systems \cite{Abughazalah_2024}. Any of these components could be vulnerable. With the existence of further offensive AI models, each vulnerability could be automatically exploited. Therefore, quantification during the step of vulnerability assessment give a means of understanding the weak points of a system.      
This paper addresses the existing research gap of quantifying the risk through AI in vulnerability assessments. It shows the current state of AI in threat modeling and introduces a new method~-- called Graph of Effort (GOE)~-- to address the danger of AI attacks in vulnerability assessments. GOE provides the exposure of the vulnerability against AI-based attacks. The method is intuitive, clear to use, and meant to be used by consumers of a given entity potentially affected by a vulnerability. GEO is based on the effort needed to create an AI model to automate the exploitation of a given vulnerability.

The remainder of the paper is structured as follow: After a brief overview of offensive AI (OAI) and AI in threat modeling in Sections \ref{sec:OAI} and \ref{sec:AIThreadModeling}, we introduce GOE and calculate the exposure of a given vulnerability of being attacked by AIs in Sections \ref{sec:quantifyAiAttackEffort} and \ref{sec:GOEsample}.
\Cref{sec:discConcl} covers a discussion on how to deal with the results of the GOE and integrate it with other vulnerability modeling systems to facilitate prioritizing vulnerabilities. Finally, implications for future work are given. 

\section{Offensive AI} \label{sec:OAI}

This Section provides an overview of how AI is offensively used for cyberattacks. According to \cite{Mirsky_2023}, offensive AI (OAI) can be grouped into two categories: (1)~attacks using AI and (2)~Adversarial Machine Learning (AML).

In the first category, OAI is used as a tool to assist adversaries in applying their attacks. Potentially, AI could resolve any task as long as it is manually done or requires the use of common thought intelligence used by humans or non-human beings. A key limiting factor for AI training are suitable datasets for training on a given task, so the model can gain experience to perform this task efficiently \cite{brundage2018malicious}.

The most common uses of OAI as a support tool according to \cite{Malatji_2024, Mirsky_2023} are:

\begin{enumerate}
    \item Prediction,
    \item Generation,
    \item Analysis,
    \item Retrieval, and
    \item Decision Making. 
\end{enumerate}

In conclusion, there are several ways to misuse AI for offensive malicious use. On the other hand, AML describes using the knowledge on how an AI model internally works to attack deployed models by organizations or other entities also via AI \cite{Thanh_2019}. Here, the AI model is attacked by an adversary to compromise its security goals, like confidentiality, integrity or availability. This can be achieved, for instance, through data poisoning, wherein adversaries introduce misleading training data to the model, or by launching a Distributed Denial of Service (DDoS) attack to degrade model performance and reduce availability \cite{Mirsky_2023}\,\cite{Malatji_2024}.

A categorization of OAI use cases may be found in \cite{schroer2024sok}. They discovered that especially in technical papers and information security briefings OAI use cases can be mostly mapped via the steps defined in MITRE ATT\&CK \cite{mitreAttack}. Meanwhile, non-technical papers mostly categorize OAI into one of the following groups:

\begin{itemize}
    \item attack in (cyber) warfare,
    \item attack on society,
    \item autonomous agents, and
    \item privacy attack.
\end{itemize}

This paper aims to prioritize the implementation of countermeasures for specific vulnerabilities, given that the effort required to execute exploits using AI is significantly lower compared to other vulnerabilities. Therefore, the main focus will be on OAI as of the first category of \cite{Mirsky_2023} and for all use cases of \cite{schroer2024sok}.

\section{AI Threat Modeling} \label{sec:AIThreadModeling}

AI is discussed in several contexts in the area of threat modeling -- this Section gives a brief overview. 

Mirsky et al.~\cite{Mirsky_2023} lists 24 offensive AI-based threats feared by organizations. Furthermore, the authors created a model to quantify the threat of AI by evaluating the harm, profit and achievability of AI-usage, as well as defeatability against it.
On the other hand, \cite{grosse2023machine} discovered that~-- at least on organizational level~-- no predictors can be found, which describe the threat exposure. However, it has to be emphasized that \cite{grosse2023machine} focused only on AML as subtopic of OAI. Therefore, further work is needed on finding threat exposure predictors on OAI. They also criticize the lack of real-world scenarios in research regarding general security of AI systems. 

Malatji and Tolah~\cite{Malatji_2024} identified a research gap in the quantification of factors leading to a better understanding of OAI usage. They specified that especially the quantification of the attackers' motivation is missing. 

Guembe et al.~\cite{Guembe_2022} found out that AI-driven cyberattacks can be done continuously throughout all steps of an attack. In addition, the authors state that current defense mechanisms will become deprecated, as AI enables more sophisticated cyberattacks and detection evasion techniques.  

Some publications focused on the creation of threat models for deployed AI systems by organizations. For instance, in \cite{Mauri_2022}, the authors created an extension of Microsoft's STRIDE threat model (STRIDE-AI) to identify vulnerabilities of deployed AI systems, approachable by consumers. STRIDE stands for Spoofing, Tampering, Repdudiation, Information Disclosure, Denial of Service and Elevation of Privilege. It aims to assist security analysts to categorize threats. The author of \cite{tete2024threat} followed a similar approach for threat modeling and risk analysis for AI systems, with a focus on Large Language Models (LLM). In his approach, the author combined the threat classification given by STRIDE with qualitative ratings for each class in the categories of Damage, Reproducibility and Exploitability (DREAD) to add the aspect of risk analysis. DREAD typically also determines the risk for Affected Users and Discoverability. However, these aspects were not analyzed in the proposed threat model in \cite{tete2024threat}. 

In \cite{von-der-assen-2024}, the authors created a web application, which assists analysts throughout the threat modeling process for AI-based IT systems. It supports especially threat identification by querying several security related databases, per instance, MITRE Atlas \cite{mitreAtlas}. By means of an underlying graph model with comparable metadata, related assets of an organization and of the databases are found and connected.

A systematic threat modeling procedure for AI software is given in \cite{kumar2024admin}. In their approach, the authors develop a process diagram for the development of AI-based software and furthermore a taxonomy of threats to AI. Based on their given process, the main idea is to sequentially go through it and analyze for all subprocesses and their respective in- and outputs potential threats to AI by means of their taxonomy. 

To the best of our knowledge, no paper deals with the creation of a threat modeling method for the exposure of general vulnerabilities against OAI, especially with a focus on being simply applicable and explainable for analysts.  

\section{Quantifying effort of AI-based attacks} \label{sec:quantifyAiAttackEffort}

\begin{figure}[!ht]
    \centering
    \begin{tikzpicture}
        \small
        \node[text width=2cm, text centered, minimum height=0.75cm, draw, thick] (recon) at (0, 0) {(1)\\Reconnaissance\vphantom{j}};
        \node[text width=2cm, text centered, minimum height=0.75cm, draw, thick] (wpn) at (1.5, -1.25) {(2)\\Weaponization};
        \node[text width=2cm, text centered, minimum height=0.75cm, draw, thick] (del) at (3, -2.5) {(3)\\Delivery};
        \node[text width=2cm, text centered, minimum height=0.75cm, draw, thick] (expl) at (4.5, -3.75) {(4)\\Exploitation};
        \draw[thick, -latex] (recon) -- (wpn);
        \draw[thick, -latex] (wpn) -- (del);
        \draw[thick, -latex] (del) -- (expl);
    \end{tikzpicture}
    \caption{Steps of the intrusion kill chain according to \cite{hutchins2011intelligence}.}
    \label{fig:hutchinsAttackChain}
\end{figure}
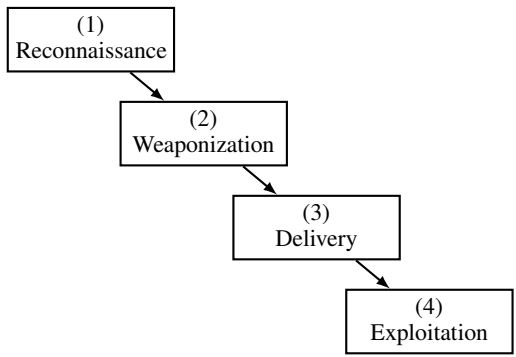

To consider and quantify the potential of AI-based attacks in vulnerability, a new method~-- graph of effort (GOE)~-- for analyzing this potential is introduced (see \Cref{sec:quantifyAiAttackEffort} for details). Foundation of the method is the modified intrusion kill chain as introduced in \cite{hutchins2011intelligence}, wherein a cyberattack consists of the four steps (1)~Reconnaissance, (2)~Weaponization, (3)~Delivery, and (4)~Exploitation as shown in \Cref{fig:hutchinsAttackChain}. Reconnaissance describes the tasks needed to find and select victims for the attack. Weaponization is the creation of a transferable package in which malicious code is hidden. Typical weaponized files were PDFs and Microsoft Office documents. The next step, Delivery, describes the transmission of the malicious package to the target network. During the final Exploitation step the, successfully transferred package activates its malicious code often targeting a vulnerability of an application or the host os itself. This kill chain was modified to align with the one used in CVSS~v4.0 \cite{First-cvss-4}. The consistent application of identical kill chains between vulnerability assessment frameworks facilitates enhanced coordination of measures during assessments.

%% TODO - Grafik ersetzen - Andreas -> done

\noindent Based upon the four kill chain steps of \cite{hutchins2011intelligence}, our metric answers the following question:

\begin{quote}
    \itshape
    What effort is needed to use offensive AI during each kill chain step?
\end{quote}

%%\textcolor{red}{discuss how those three questions cover all the space.}

\noindent Our metric calculates a score for each respective step $i,\;1\leq i\leq 4,$ in the kill chain, see \Cref{eq:AIES}:

\begin{equation}
    \text{score}_{(i)} \in\lbrace 0,\dots,3\rbrace.
    \label{eq:AIES}
\end{equation}

%%\textcolor{red}{Entscheidung bewusst für binäre Entscheidungen pro Stufe, denn mehr Bewertungsmöglichkeiten sorgen nicht dafür, dass die Bewertung objektiver wird. Beispiel: STRIDE/DREAD, Bewertung jeweils schwach/mittel/stark  -> sehr viel mehr Subjektivität bei Abwägung z. B. zwischen "mittel" und "stark"}

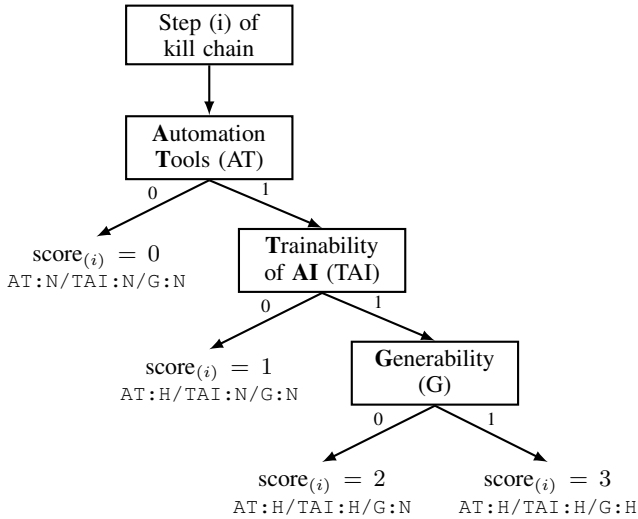
\begin{figure}[ht!]
    \centering
    \small
    \begin{tikzpicture}
        \node[text width=2cm, minimum height=0.5cm, text centered, draw, thick] (kc) at (0, 0) {Step (i) of\\kill chain};
        \node[text width=2cm, minimum height=0.5cm, text centered, draw, thick] (at) at (0, -1.5) {\textbf{A}utomation \textbf{T}ools (AT)};
        \node[text width=2cm, minimum height=0.5cm, text centered, draw, thick] (tai) at (1.5, -3) {\textbf{T}rainability of \textbf{AI} (TAI)};
        \node[text width=2cm, minimum height=0.5cm, text centered, draw, thick] (g) at (3, -4.5) {\textbf{G}enerability (G)};
        \draw[thick, -latex] (kc) -- (at);
        \draw[thick, -latex] (at.south) -- node[above]{\scriptsize 0} ++(-1.5, -0.75) node[below, text width=2.5cm, text centered]{$\text{score}_{(i)}=0$\\\footnotesize\texttt{AT:N/TAI:N/G:N}};
        \draw[thick, -latex] (at.south) -- node[above]{\scriptsize 1} (tai.north);
        \draw[thick, -latex] (tai.south) -- node[above]{\scriptsize 0} ++(-1.5, -0.75) node[below, text width=2.5cm, text centered]{$\text{score}_{(i)}=1$\\\footnotesize\texttt{AT:H/TAI:N/G:N}};
        \draw[thick, -latex] (tai.south) -- node[above]{\scriptsize 1} (g.north);
        \draw[thick, -latex] (g.south) -- node[above]{\scriptsize 0} ++(-1.5, -0.75) node[below, text width=2.5cm, text centered]{$\text{score}_{(i)}=2$\\\footnotesize\texttt{AT:H/TAI:H/G:N}};
        \draw[thick, -latex] (g.south) -- node[above]{\scriptsize 1} ++(1.5, -0.75) node[below, text width=2.5cm, text centered]{$\text{score}_{(i)}=3$\\\footnotesize\texttt{AT:H/TAI:H/G:H}};
    \end{tikzpicture}
    \caption{Visualization of the GOE to calculate the effort needed to use AI for an attack step in the intrusion kill chain according to \cite{hutchins2011intelligence}.}
    \label{fig:treeOfEffort}
\end{figure}

%% TODO - "weights" ersetzen mit was Anderem ==> done
\noindent The GOE is visualized in \Cref{fig:treeOfEffort} as a binary tree, in which each node describes an advanced level of effort for AI usage the attacker has to muster from top to bottom. The idea of the GOE is to obtain a score based on the answers to three questions: Are there \dots
\begin{itemize}
    \item[\dots] ready to use models or AI-based tools,
    \item[\dots] datasets to train a suitable model, or
    \item[\dots] automatisms available to quickly generate suitable data to train an AI model? 
\end{itemize}

%% TODO - Das sollte doch später im Text verwendet werden.
The purpose of the questions is to address all relevant aspects of using or training an AI system. The primary assumption is that the simplest way to utilize AI is through ready-to-use systems equipped with a well-designed Graphical User Interface (GUI). In the absence of a GUI, pretrained models can also be quickly deployed.

It is assumed that the adversary possesses broad knowledge of AI and has access to all necessary resources for deployment. Therefore, if a ready-to-use AI system is not available, the adversary could train their own model. This process requires data, which is the most critical asset in training an AI. Without data, AI solutions cannot be developed. Consequently, for the adversary to create an AI system, they must have suitable training data. If such data is not available, the adversary must generate their own dataset. However, given that the provider of a CVE (CVE) is interested in promptly addressing the issue, it is further assumed that the adversary often lacks the time to manually create a high-quality dataset. Instead, they must rely on automated methods to generate data as quickly as possible.

The questions do not cover the detailed sub-aspects related to AI tools and datasets. For instance, AI tools can differ significantly; some are open-source, while others require specialized expertise. The same variability applies to datasets. This design choice is intentional. By avoiding excessive detail, the GOE ensures objectivity. The questions, though simple in design, allow for concise answers, limited to "yes" or "no," thereby minimizing subjective interpretations. For example, if a tool is deemed to require expertise, this perception might initially seem subjective, as individuals have varying definitions of expertise. Other threat assessment systems, such as STRIDE, allow users to select from predefined categorical values like "low/medium/high." However, GOE has intentionally omitted such values to maintain objectivity.

As already indicated in \Cref{eq:AIES}, the scores range from $0$ (no or low) to $3$ (high) to describe the attackers effort; scores depicted on the left-hand side in \Cref{fig:treeOfEffort} are always smaller than those on the right-hand side of each level. Starting at the top for each step of the kill chain, the scores and, therefore, the attacker's effort are increasing downwards. The process is stopped when one of the leaves is reached, which are indicated by the term $\text{score}_{(i)}$.\par 
The first question that has to be answered is whether AI-based tools, AI models that are ready to use, or AI-based automatisms already exist. If this the case, then this represents the least possible effort for the attacker ($\text{score}_{(i)}=0$) because they may directly use these tools to automate their attack. In order to be able to use GOE together with CVSS, we propose a corresponding extension of the vector string. Since GOE only uses binary decisions, we set $0$ to \texttt{N} and $1$ to \texttt{H} in order to correspond to the symbols used in CVSS. The first category ``\textbf{A}utomation \textbf{T}ools'', \texttt{AT:0} or \texttt{AT:1}, can thus be noted for the top level of the GOE according to the decision left (0) or right (1) in the binary tree.\par 
Given \texttt{AT:1}, so there are neither AI-based tools, ready to use models nor AI-based automatisms available, the next question is whether ready to use datasets or even complete training setups exist which the attacker may use to generate their own AI models. Again, this question may be answered with ``yes'' (left) which results in the substring \texttt{TAI:0} (for \textbf{T}rainability of \textbf{AI}) and $\text{score}_{(i)}=1$. If no such datasets or training setups exist (indicated by \texttt{TAI:1}), the final criterion, \textbf{G}enerability, has to be assessed: Are there APIs or any other tools that enable the automatic creation of data sets to create an AI model? \texttt{G:0} means ``yes'', resulting in $\text{score}_{(i)}=2$, whereas \texttt{G:1} indicates the greatest possible effort for the adversary ($\text{score}_{(i)}=3$).

If the remaining categories, in the order \texttt{AT}, \texttt{TAI}, and \texttt{G}, are set to \texttt{N} (which is equivalent to $0$), after a leaf has been reached, then the score for step~$i$ of the kill chain may also be calculated as the sum of the three categories (\texttt{H} is equivalent to $1$):
\begin{equation}
    \text{score}_{(i)} = \mathtt{AT} + \mathtt{TAI} + \mathtt{G}\quad
\end{equation}

The proposed threat model determines the AI-based threat with an intuitive approach, which is clearly understandable and easy to visualize. Additionally, GOE is flexible in its usage. Depending on the specific vulnerability or the priorities of security analysts, some of the kill chain attack steps may easily be skipped.

To calculate the overall score $\text{GOE}(v)$ for a given vulnerability $v$, \Cref{eq:finalAIES} is used:

%% TODO - Variablennamen umändern
\begin{equation}
    \text{GOE}(v) = \min_i\left\lbrace\text{score}_{(i)}\right\rbrace
    \label{eq:finalAIES}
\end{equation}

Using the minimum score of all steps in the kill chain as the overall score seems most reasonable and intuitive because if one of the steps is easily exploitable through AI, then it effects the exposure of the vulnerability as a whole. But \Cref{eq:finalAIES} may also be adapted if the analyst prefers, e.g., a weighted average of the scores of the four steps. Additionally, we would like to mention that the assessment of steps of the kill chain may be skipped, if these are of no interest for the analyst. If it is nevertheless desired to use \Cref{eq:finalAIES}, the score of the steps not taken into account can be set to infinity ($\infty$), for example.

The introduction of GOE based on CVSS is intended merely as an example. The universality of GOE allows it to be combined with any method of threat and risk analysis if it is necessary or desired to express whether a vulnerability should be assessed differently due to the availability of AI models. As already indicated, direct integration of GOE in CVSS v4.0 is possible via the “Automatable” criterion in the optional Supplemental Metric Group~\cite{First-cvss-4}. If CVSS is combined with GOE, one could assume that if a given vulnerability is completely automatable, indicated via a positive value of the ``Automatable'' criterion. 

To achieve clear and intuitive applicability of the model, some assumptions are made. The adversary is assumed to have unlimited resources and knowledge regarding the considered vulnerability as well as skills in managing and training AI solutions. This assumption ensures that an attacker is not underestimated. Moreover, the proposed GOE does not allow higher scores than $3$ (if there is no method of generating data automatically, \texttt{AT:H/TAI:H/G:H}). One could now argue that data could be created manually since the adversary according to our first assumption has unlimited resources to do so. However, there is only a small amount of time to create AI-based solutions for vulnerability exploitation because affected service providers are interested to deliver fixes or at least workarounds to their customers or users as soon as possible. Furthermore, if not much data is needed to create a successful AI model, it may be assumed that rule-based automation can easily be created~-- which is covered by our proposed AT criterion leading to $\text{score}_i = 0$ and, therefore, $\text{GOE}(v) = 0$.   

\section{Example Scoring - CVE-2025-1156} \label{sec:GOEsample}

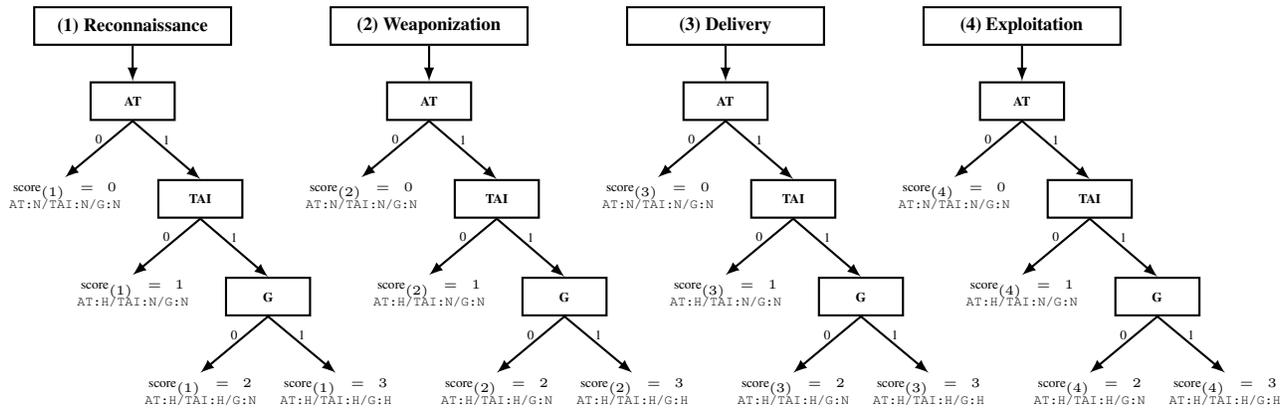
\begin{figure*}[t!]
    \centering
    \tiny
    \begin{tikzpicture}
        \node[text width=2.5cm, minimum height=0.5cm, text centered, draw, thick] (kc) at (0, 0) {\scriptsize\textbf{(1) Reconnaissance}};
        \node[text width=1cm, minimum height=0.5cm, text centered, draw, thick] (at) at (0, -1) {\textbf{AT}};
        \node[text width=1cm, minimum height=0.5cm, text centered, draw, thick] (tai) at (0.9, -2.3) {\textbf{TAI}};
        \node[text width=1cm, minimum height=0.5cm, text centered, draw, thick] (g) at (1.8, -3.6) {\textbf{G}};
        \draw[thick, -latex] (kc) -- (at);
        \draw[thick, -latex, opacity=0.3] (at.south) -- node[above, opacity=0.3]{0} ++(-0.9, -0.75) node[below, text width=2.5cm, text centered, opacity=0.3]{$\text{score}_{(1)}=0$\\\texttt{AT:N/TAI:N/G:N}};
        \draw[thick, -latex] (at.south) -- node[above]{1} (tai.north);
        \draw[thick, -latex, opacity=0.3] (tai.south) -- node[above, opacity=0.3]{0} ++(-0.9, -0.75) node[below, text width=2.5cm, text centered, opacity=0.3]{$\text{score}_{(1)}=1$\\\texttt{AT:H/TAI:N/G:N}};
        \draw[thick, -latex] (tai.south) -- node[above]{1} (g.north);
        \draw[thick, -latex] (g.south) -- node[above]{0} ++(-0.9, -0.75) node[below, text width=2.5cm, text centered]{$\text{score}_{(1)}=2$\\\texttt{AT:H/TAI:H/G:N}};
        \draw[thick, -latex, opacity=0.3] (g.south) -- node[above, opacity=0.3]{1} ++(0.9, -0.75) node[below, text width=2.5cm, text centered, opacity=0.3]{$\text{score}_{(1)}=3$\\\texttt{AT:H/TAI:H/G:H}};
    \end{tikzpicture}
    \hspace{-2.4cm}
    \begin{tikzpicture}
        \node[text width=2.5cm, minimum height=0.5cm, text centered, draw, thick] (kc) at (0, 0) {\scriptsize\textbf{(2) Weaponization}};
        \node[text width=1cm, minimum height=0.5cm, text centered, draw, thick] (at) at (0, -1) {\textbf{AT}};
        \node[text width=1cm, minimum height=0.5cm, text centered, draw, thick, opacity=0.3] (tai) at (0.9, -2.3) {\textbf{TAI}};
        \node[text width=1cm, minimum height=0.5cm, text centered, draw, thick, opacity=0.3] (g) at (1.8, -3.6) {\textbf{G}};
        \draw[thick, -latex] (kc) -- (at);
        \draw[thick, -latex] (at.south) -- node[above]{0} ++(-0.9, -0.75) node[below, text width=2.5cm, text centered]{$\text{score}_{(2)}=0$\\\texttt{AT:N/TAI:N/G:N}};
        \draw[thick, -latex, opacity=0.3] (at.south) -- node[above, opacity=0.3]{1} (tai.north);
        \draw[thick, -latex, opacity=0.3] (tai.south) -- node[above, opacity=0.3]{0} ++(-0.9, -0.75) node[below, text width=2.5cm, text centered]{$\text{score}_{(2)}=1$\\\texttt{AT:H/TAI:N/G:N}};
        \draw[thick, -latex, opacity=0.3] (tai.south) -- node[above, opacity=0.3]{1} (g.north);
        \draw[thick, -latex, opacity=0.3] (g.south) -- node[above, opacity=0.3]{0} ++(-0.9, -0.75) node[below, text width=2.5cm, text centered, opacity=0.3]{$\text{score}_{(2)}=2$\\\texttt{AT:H/TAI:H/G:N}};
        \draw[thick, -latex, opacity=0.3] (g.south) -- node[above]{1} ++(0.9, -0.75) node[below, text width=2.5cm, text centered]{$\text{score}_{(2)}=3$\\\texttt{AT:H/TAI:H/G:H}};
    \end{tikzpicture}
    \hspace{-2.4cm}
    \begin{tikzpicture}
        \node[text width=2.5cm, minimum height=0.5cm, text centered, draw, thick] (kc) at (0, 0) {\scriptsize\textbf{(3) Delivery}};
        \node[text width=1cm, minimum height=0.5cm, text centered, draw, thick] (at) at (0, -1) {\textbf{AT}};
        \node[text width=1cm, minimum height=0.5cm, text centered, draw, thick, opacity=0.3] (tai) at (0.9, -2.3) {\textbf{TAI}};
        \node[text width=1cm, minimum height=0.5cm, text centered, draw, thick, opacity=0.3] (g) at (1.8, -3.6) {\textbf{G}};
        \draw[thick, -latex] (kc) -- (at);
        \draw[thick, -latex] (at.south) -- node[above]{0} ++(-0.9, -0.75) node[below, text width=2.5cm, text centered]{$\text{score}_{(3)}=0$\\\texttt{AT:N/TAI:N/G:N}};
        \draw[thick, -latex, opacity=0.3] (at.south) -- node[above, opacity=0.3]{1} (tai.north);
        \draw[thick, -latex, opacity=0.3] (tai.south) -- node[above, opacity=0.3]{0} ++(-0.9, -0.75) node[below, text width=2.5cm, text centered]{$\text{score}_{(3)}=1$\\\texttt{AT:H/TAI:N/G:N}};
        \draw[thick, -latex, opacity=0.3] (tai.south) -- node[above, opacity=0.3]{1} (g.north);
        \draw[thick, -latex, opacity=0.3] (g.south) -- node[above, opacity=0.3]{0} ++(-0.9, -0.75) node[below, text width=2.5cm, text centered, opacity=0.3]{$\text{score}_{(3)}=2$\\\texttt{AT:H/TAI:H/G:N}};
        \draw[thick, -latex, opacity=0.3] (g.south) -- node[above]{1} ++(0.9, -0.75) node[below, text width=2.5cm, text centered]{$\text{score}_{(3)}=3$\\\texttt{AT:H/TAI:H/G:H}};
    \end{tikzpicture}
    \hspace{-2.4cm}
    \begin{tikzpicture}
        \node[text width=2.5cm, minimum height=0.5cm, text centered, draw, thick, opacity=0.3] (kc) at (0, 0) {\scriptsize\textbf{(4) Exploitation}};
        \node[text width=1cm, minimum height=0.5cm, text centered, draw, thick, opacity=0.3] (at) at (0, -1) {\textbf{AT}};
        \node[text width=1cm, minimum height=0.5cm, text centered, draw, thick, opacity=0.3] (tai) at (0.9, -2.3) {\textbf{TAI}};
        \node[text width=1cm, minimum height=0.5cm, text centered, draw, thick, opacity=0.3] (g) at (1.8, -3.6) {\textbf{G}};
        \draw[thick, -latex, opacity=0.3] (kc) -- (at);
        \draw[thick, -latex, opacity=0.3] (at.south) -- node[above, opacity=0.3]{0} ++(-0.9, -0.75) node[below, text width=2.5cm, text centered, opacity=0.3]{$\text{score}_{(4)}=0$\\\texttt{AT:N/TAI:N/G:N}};
        \draw[thick, -latex, opacity=0.3] (at.south) -- node[above, opacity=0.3]{1} (tai.north);
        \draw[thick, -latex, opacity=0.3] (tai.south) -- node[above, opacity=0.3]{0} ++(-0.9, -0.75) node[below, text width=2.5cm, text centered]{$\text{score}_{(4)}=1$\\\texttt{AT:H/TAI:N/G:N}};
        \draw[thick, -latex, opacity=0.3] (tai.south) -- node[above, opacity=0.3]{1} (g.north);
        \draw[thick, -latex, opacity=0.3] (g.south) -- node[above, opacity=0.3]{0} ++(-0.9, -0.75) node[below, text width=2.5cm, text centered, opacity=0.3]{$\text{score}_{(4)}=2$\\\texttt{AT:H/TAI:H/G:N}};
        \draw[thick, -latex, opacity=0.3] (g.south) -- node[above]{1} ++(0.9, -0.75) node[below, text width=2.5cm, text centered]{$\text{score}_{(4)}=3$\\\texttt{AT:H/TAI:H/G:H}};
    \end{tikzpicture}
    \caption{Visualization of the GOE for the known vulnerability CVE-2025-1156 listed in the National Vulnerability Database (NVD), showing the effort needed to use AI in each step of the intrusion kill chain. Given the flexibility of GOE, step (4) of the kill chain is skipped in this case. The overall score is $\text{GOE}=0$, corresponding to a low effort for exploitation by AI.}
    \label{fig:treeOfEffort_Samp1}
\end{figure*}

After introducing GOE as an intuitive approach, this Section provides an example modeling for a known vulnerability, namely CVE-2025-1156 listed in the National Vulnerability Database (NVD)\cite{CVE-2025-1156}:
\begin{quote}
    ``A vulnerability has been found in Pix Software Vivaz 6.0.10 and classified as critical. This vulnerability affects unknown code of the file \texttt
    {/servlet?act=login}. The manipulation of the argument usuario leads to sql injection. The attack can be initiated remotely. The exploit has been disclosed to the public and may be used. The vendor was contacted early about this disclosure but did not respond in any way.''
\end{quote}
Thus, the chosen vulnerability describes an improperly sanitized user input while requesting a given HTTP endpoint for authentication, making it vulnerable to Structured Query Language (SQL) injection. The complexity to launch such an attack is low as only websites using this specific service-desk software need to be found to do the injection \cite{CVE-2024-30384}. \par 
It is worth noticing, that GOE may be used to analyze any given vulnerability. We specifically chose this vulnerability as an example because:
\begin{itemize}
    \item it is a relatively new entry in the NVD,
    \item the vulnerability is relevant for cloud services, and
    \item we can use it to demonstrate that GOE can be used in conjunction with CVSS v4.0 as well as with earlier versions, e.g. CVSS v3.x.
\end{itemize} 
According to the NVD, the vectors for CVE-2025-1156 are as follows:
\begin{quote}
     \texttt{CVSS:3.1/AV:N/AC:L/PR:N/UI:N/S:U/\\C:L/I:L/A:L}\par
     and\par
     \texttt{CVSS:4.0/AV:N/AC:L/AT:N/PR:N/UI:N/\\VC:L/VI:L/VA:L/SC:N/SI:N/SA:N},
\end{quote}
resulting in a base score of $7.3$ (criticality ``high'') for CVSS v3.x and a CVSS-B score of $6.9$ (criticality ``medium'') for CVSS v4.0.
We now evaluate the vulnerability using the method described above and calculate the associated GOE, see also Figure \ref{fig:treeOfEffort_Samp1} for a visualization of the GOE for every single step of the kill chain.\par 

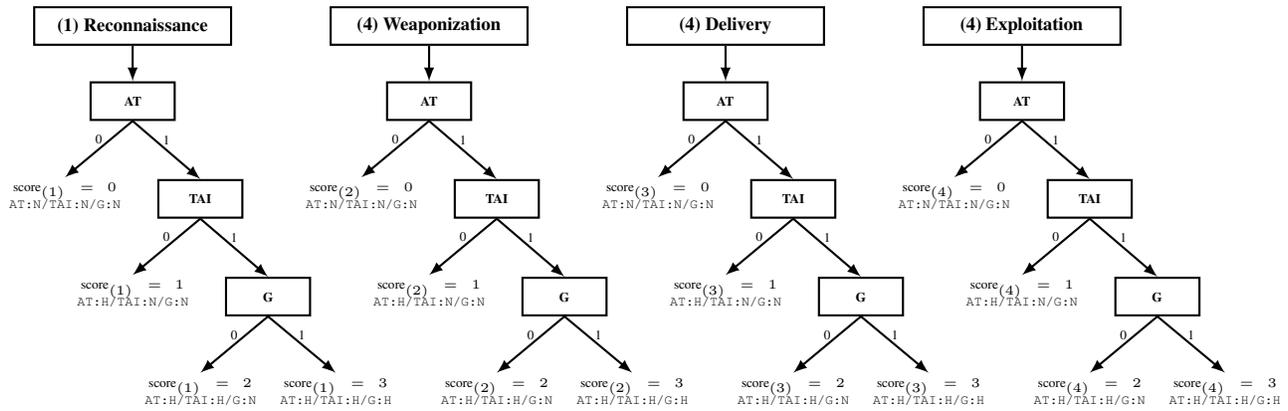
\begin{figure*}[t!]
    \centering
    \tiny
    \begin{tikzpicture}
        \node[text width=2.5cm, minimum height=0.5cm, text centered, draw, thick] (kc) at (0, 0) {\scriptsize\textbf{(1) Reconnaissance}};
        \node[text width=1cm, minimum height=0.5cm, text centered, draw, thick] (at) at (0, -1) {\textbf{AT}};
        \node[text width=1cm, minimum height=0.5cm, text centered, draw, thick] (tai) at (0.9, -2.3) {\textbf{TAI}};
        \node[text width=1cm, minimum height=0.5cm, text centered, draw, thick] (g) at (1.8, -3.6) {\textbf{G}};
        \draw[thick, -latex] (kc) -- (at);
        \draw[thick, -latex, opacity=0.3] (at.south) -- node[above, opacity=0.3]{0} ++(-0.9, -0.75) node[below, text width=2.5cm, text centered, opacity=0.3]{$\text{score}_{(1)}=0$\\\texttt{AT:N/TAI:N/G:N}};
        \draw[thick, -latex] (at.south) -- node[above]{1} (tai.north);
        \draw[thick, -latex, opacity=0.3] (tai.south) -- node[above, opacity=0.3]{0} ++(-0.9, -0.75) node[below, text width=2.5cm, text centered, opacity=0.3]{$\text{score}_{(1)}=1$\\\texttt{AT:H/TAI:N/G:N}};
        \draw[thick, -latex] (tai.south) -- node[above]{1} (g.north);
        \draw[thick, -latex, opacity=0.3] (g.south) -- node[above, opacity=0.3]{0} ++(-0.9, -0.75) node[below, text width=2.5cm, text centered, opacity=0.3]{$\text{score}_{(1)}=2$\\\texttt{AT:H/TAI:H/G:N}};
        \draw[thick, -latex] (g.south) -- node[above]{1} ++(0.9, -0.75) node[below, text width=2.5cm, text centered]{$\text{score}_{(1)}=3$\\\texttt{AT:H/TAI:H/G:H}};
    \end{tikzpicture}
    \hspace{-2.4cm}
    \begin{tikzpicture}
        \node[text width=2.5cm, minimum height=0.5cm, text centered, draw, thick, opacity=0.3] (kc) at (0, 0) {\scriptsize\textbf{(4) Weaponization}};
        \node[text width=1cm, minimum height=0.5cm, text centered, draw, thick, opacity=0.3] (at) at (0, -1) {\textbf{AT}};
        \node[text width=1cm, minimum height=0.5cm, text centered, draw, thick, opacity=0.3] (tai) at (0.9, -2.3) {\textbf{TAI}};
        \node[text width=1cm, minimum height=0.5cm, text centered, draw, thick, opacity=0.3] (g) at (1.8, -3.6) {\textbf{G}};
        \draw[thick, -latex, opacity=0.3] (kc) -- (at);
        \draw[thick, -latex, opacity=0.3] (at.south) -- node[above, opacity=0.3]{0} ++(-0.9, -0.75) node[below, text width=2.5cm, text centered, opacity=0.3]{$\text{score}_{(2)}=0$\\\texttt{AT:N/TAI:N/G:N}};
        \draw[thick, -latex, opacity=0.3] (at.south) -- node[above, opacity=0.3]{1} (tai.north);
        \draw[thick, -latex, opacity=0.3] (tai.south) -- node[above, opacity=0.3]{0} ++(-0.9, -0.75) node[below, text width=2.5cm, text centered]{$\text{score}_{(2)}=1$\\\texttt{AT:H/TAI:N/G:N}};
        \draw[thick, -latex, opacity=0.3] (tai.south) -- node[above, opacity=0.3]{1} (g.north);
        \draw[thick, -latex, opacity=0.3] (g.south) -- node[above, opacity=0.3]{0} ++(-0.9, -0.75) node[below, text width=2.5cm, text centered, opacity=0.3]{$\text{score}_{(2)}=2$\\\texttt{AT:H/TAI:H/G:N}};
        \draw[thick, -latex, opacity=0.3] (g.south) -- node[above]{1} ++(0.9, -0.75) node[below, text width=2.5cm, text centered]{$\text{score}_{(2)}=3$\\\texttt{AT:H/TAI:H/G:H}};
    \end{tikzpicture}
    \hspace{-2.4cm}
    \begin{tikzpicture}
        \node[text width=2.5cm, minimum height=0.5cm, text centered, draw, thick, opacity=0.3] (kc) at (0, 0) {\scriptsize\textbf{(4) Delivery}};
        \node[text width=1cm, minimum height=0.5cm, text centered, draw, thick, opacity=0.3] (at) at (0, -1) {\textbf{AT}};
        \node[text width=1cm, minimum height=0.5cm, text centered, draw, thick, opacity=0.3] (tai) at (0.9, -2.3) {\textbf{TAI}};
        \node[text width=1cm, minimum height=0.5cm, text centered, draw, thick, opacity=0.3] (g) at (1.8, -3.6) {\textbf{G}};
        \draw[thick, -latex, opacity=0.3] (kc) -- (at);
        \draw[thick, -latex, opacity=0.3] (at.south) -- node[above, opacity=0.3]{0} ++(-0.9, -0.75) node[below, text width=2.5cm, text centered, opacity=0.3]{$\text{score}_{(3)}=0$\\\texttt{AT:N/TAI:N/G:N}};
        \draw[thick, -latex, opacity=0.3] (at.south) -- node[above, opacity=0.3]{1} (tai.north);
        \draw[thick, -latex, opacity=0.3] (tai.south) -- node[above, opacity=0.3]{0} ++(-0.9, -0.75) node[below, text width=2.5cm, text centered]{$\text{score}_{(3)}=1$\\\texttt{AT:H/TAI:N/G:N}};
        \draw[thick, -latex, opacity=0.3] (tai.south) -- node[above, opacity=0.3]{1} (g.north);
        \draw[thick, -latex, opacity=0.3] (g.south) -- node[above, opacity=0.3]{0} ++(-0.9, -0.75) node[below, text width=2.5cm, text centered, opacity=0.3]{$\text{score}_{(3)}=2$\\\texttt{AT:H/TAI:H/G:N}};
        \draw[thick, -latex, opacity=0.3] (g.south) -- node[above]{1} ++(0.9, -0.75) node[below, text width=2.5cm, text centered]{$\text{score}_{(3)}=3$\\\texttt{AT:H/TAI:H/G:H}};
    \end{tikzpicture}
    \hspace{-2.4cm}
    \begin{tikzpicture}
        \node[text width=2.5cm, minimum height=0.5cm, text centered, draw, thick, opacity=0.3] (kc) at (0, 0) {\scriptsize\textbf{(4) Exploitation}};
        \node[text width=1cm, minimum height=0.5cm, text centered, draw, thick, opacity=0.3] (at) at (0, -1) {\textbf{AT}};
        \node[text width=1cm, minimum height=0.5cm, text centered, draw, thick, opacity=0.3] (tai) at (0.9, -2.3) {\textbf{TAI}};
        \node[text width=1cm, minimum height=0.5cm, text centered, draw, thick, opacity=0.3] (g) at (1.8, -3.6) {\textbf{G}};
        \draw[thick, -latex, opacity=0.3] (kc) -- (at);
        \draw[thick, -latex, opacity=0.3] (at.south) -- node[above, opacity=0.3]{0} ++(-0.9, -0.75) node[below, text width=2.5cm, text centered, opacity=0.3]{$\text{score}_{(4)}=0$\\\texttt{AT:N/TAI:N/G:N}};
        \draw[thick, -latex, opacity=0.3] (at.south) -- node[above, opacity=0.3]{1} (tai.north);
        \draw[thick, -latex, opacity=0.3] (tai.south) -- node[above, opacity=0.3]{0} ++(-0.9, -0.75) node[below, text width=2.5cm, text centered]{$\text{score}_{(4)}=1$\\\texttt{AT:H/TAI:N/G:N}};
        \draw[thick, -latex, opacity=0.3] (tai.south) -- node[above, opacity=0.3]{1} (g.north);
        \draw[thick, -latex, opacity=0.3] (g.south) -- node[above, opacity=0.3]{0} ++(-0.9, -0.75) node[below, text width=2.5cm, text centered, opacity=0.3]{$\text{score}_{(4)}=2$\\\texttt{AT:H/TAI:H/G:N}};
        \draw[thick, -latex, opacity=0.3] (g.south) -- node[above]{1} ++(0.9, -0.75) node[below, text width=2.5cm, text centered]{$\text{score}_{(4)}=3$\\\texttt{AT:H/TAI:H/G:H}};
    \end{tikzpicture}
    \caption{Visualization of the GOE for the vulnerability CVE-2024-30384, showing the effort needed to use AI in each step of the intrusion kill chain. Given the flexibility of GOE, steps (2-4) of the kill chain are skipped in this case. The overall score is $\text{GOE}=3$, corresponding to a high effort for exploitation by AI and demonstrating the GoE can have values other than 0.}
    \label{fig:treeOfEffort_Samp2}
\end{figure*}

\vspace{0.1cm}\noindent\textbf{(1)  Reconnaissance.} An AI model for reconnaissance can detect potential victims (websites) using the given service-desk software. To the best of our knowledge, there are no tools or AI models which could be used to find suitable victims, using the given software. A ready-to-use dataset to train own models is also not available. However, automatically generating training data is possible as it requires only a webcrawler visiting several websites. This crawler has to be capable of finding login pages of the vendor of this service-desk software. This can be done via image recognition or via parsing the Document Object Model (DOM) of the website and comparing the id and class names of the respective HTML elements, for instance. So, the GOE sub-vector looks as follows:

\begin{quote}
    \texttt{AT:H/TAI:H/G:N},
\end{quote}

\noindent resulting in a level score of $\text{score}_{(1)} = 2$, indicating that AI-based attacks are feasible but are connected with a higher score. For this task, models have to be specifically created by the adversary.\par 

\vspace{0.1cm}\noindent\textbf{(2) Weaponization.} A weaponization AI model for this vulnerability needs to create a malicious HTTP request, incorporating the vulnerable query parameter. Basically, this can be done via string interpolation and, therefore, does not need an AI model at all. However, Large Language Models (LLMs) can be used to create malicious HTTP requests. The GOE vector looks as follows:

\begin{quote}
    \texttt{AT:N/TAI:N/G:N}
\end{quote}

\noindent and $\text{score}_{(2)} = 0$ indicates, that no effort is needed to incorporate AI to automate the generation of malicious HTTP requests.\par 

\vspace{0.1cm}\noindent\textbf{(3) Delivery.} An AI model for step (3) of the kill chain needs to transport the string used for the SQL injection to the victim network. This is done via a simple HTTP request. Therefore, no AI-based automation is needed. However, AI-based tools, such as LLMs, may be incorporated here which can automate or assist in creating the HTTP request. Therefore, we calculate  $\text{score}_{(3)} = 0$, resulting in the same sub-vector as in the previous step:

\begin{quote}
    \texttt{AT:N/TAI:N/G:N}
\end{quote}

\noindent It is important to highlight that AI models potentially can be used to evade detection mechanisms \cite{Thanh_2019} for misuse, such as SQL injections. One might thus argue that multiple AI models could be used in this step. In that case, our recommendation is to calculate the GOE sub-vector for each AI model and then use the one with the lowest score to describe it for the respective step since GOE is always meant as a worst case analysis.\par 

\vspace{0.1cm}\noindent\textbf{(4) Exploitation.} An AI model for exploitation should typically prepare or activate the malicious code after it has been successfully delivered. However, since the malicious code gets inserted into the backend, no further doings are required here for activation. Therefore, this step is skipped. This also highlights the flexible usage of the GOE, since steps can easily be skipped if not needed without effecting the score calculation.

%%\textcolor{red}{Zweites Beispiel, bei dem GOE != 0 gilt, genauso durchdeklinieren wie erstes Beispiel}

In conclusion, CVE-2025-1156 gets an overall GOE score of
$$\text{GOE}(\text{CVE-2025-1156})=\min\lbrace2,0,0,\infty\rbrace= 0.$$ 

This score means that using OAI to assist in this attack is connected with low effort. As OAI allows for a higher efficiency, scaling and attack automation, this CVE could be more exploited via AI than CVEs with a higher GOE rating. In a real life vulnerability assessment environment, this CVE should be higher prioritized regarding the implementation of AI countermeasures such as, f.e. request rate limiting systems or captchas. Analysts can further combine the high CVSS base score with the low GOE score and could raise the criticality. Thus, to raise the prioritization of dealing with the CVE and enhancing their arguments with the information of the GOE.

\section{Example Scoring - CVE-2024-30384}

In this Section, we do another example scoring for CVE-2024-30384 (see Figure \ref{fig:treeOfEffort_Samp2} for a visualization of the GOE for every
single step of the kill chain). The vulnerability is described as follow:

\begin{quote}
    ``An Improper Check for Unusual or Exceptional Conditions vulnerability in the Packet Forwarding Engine (PFE) of Juniper Networks Junos OS on EX4300 Series allows a locally authenticated attacker with low privileges to cause a Denial-of-Service (DoS). If a specific CLI command is issued, a PFE crash will occur. This will cause traffic forwarding to be interrupted until the system self-recovers. This issue affects Junos OS: All versions before 20.4R3-S10, 21.2 versions before 21.2R3-S7, 21.4 versions before 21.4R3-S6.''
\end{quote}

Hence, the vulnerability is the usage of a preinstalled CLI application in Junos OS on their enterprise switch EX4300 \cite{Juniper-Switch}. Junos OS is the operating system based on FreeBSD or Linux of the vendor Juniper Network Devices. The CVSS v3.1 and v4.0 vectors are as follow:

\begin{quote}
     \texttt{CVSS:3.1/AV:L/AC:L/PR:L/UI:N/S:U\\/C:N/I:N/A:H}\par 
     and\par
     \texttt{CVSS:4.0/AV:L/AC:L/AT:N/PR:L/UI:N\\/VC:N/VI:N/VA:H/SC:N/SI:N/SA:L}\par
    \end{quote}

\noindent The base score is 5.5 for v3.1 (criticality “Medium”) and 6.8 f   or v4.0 (criticality “Medium”).
This vulnerability is chosen as second example because it

\begin{itemize}
    \item needs local access to be exploited,
    \item shows that AI can sophisticate attacks and therefore, should not always need to be deployed, and
    \item demonstrates that GOE can have values other than 0.
\end{itemize}

\vspace{0.1cm}\noindent\textbf{(1)  Reconnaissance.} To find suitable victims, an adversary needs to find enterprises which use these switches. The switch can be utilized to work as layer 2 (data link layer) or layer 3 (network layer) device. Hence, not all devices can be found via network exploration. Furthermore, network switches are not open to the internet. Thus, an adversary needs to have physical access or contacts within the victim organization to know if suitable devices are existent within. Acquiring the access or even at least the information is a highly individual process. To the best of our knowledge no tools can assist here. Training a model is insufficient too. Because the process is different each time, we cannot even say what kind of task the model should be trained on, and with what data, to help detect this CVE. We therefore give a reconnaissance score of 3. The GOE sub-vector looks as follows:

\begin{quote}
    \texttt{AT:H/TAI:H/G:H},
\end{quote}

\vspace{0.1cm}\noindent\textbf{(2) Weaponization, (3) Delivery.} Since the "weapon" in the vulnerability is the pre-installed software itself weaponization and delivery can be skipped.

\vspace{0.1cm}\noindent\textbf{(4)  Exploitation.} For the exploitation, an uncertain CLI but preinstalled command needs to be entered. The low attack complexity in the CVSS vectors indicate that it's not a sophisticated attack where timing or patterns needs to be known, even if it's only an assumption. Training an AI seems unreasonable here, therefore this step is also skipped. It would be more suited to use Robotic Process Automation to run the malicious command.

In conclusion, CVE-2024-30384 gets a GOE score of 3 with the following vector: $$\text{GOE}(\text{CVE-2024-30384})=\min\lbrace3,\infty,\infty,\infty\rbrace= 3$$ 
It is unlikely that AI is used in any form to assist in exploiting this CVE. In a hypothetical setting in which CVEs were assessed and prioritized only using GOE, this CVE could be ignored.

\section{Conclusion and Future Work} \label{sec:discConcl}
With GOE, an intuitive yet effective method is provided to assess the exposure of AI, based on the effort required for the attacker. GOE offers a flexible approach to quantify and provide a simplistic explanation of AI usage in vulnerability exploitation. After rating vulnerabilities by means of GOE, analysts and other stakeholders can prioritize which vulnerabilities should be assessed first. By combining GOE with well-known and established vulnerability assessment systems such as CVSS, a comprehensive analysis of vulnerabilities can be achieved and more detailed ratings can be done. The current GOE implementation consists of four possible rating values. Therefore, the likelihood is high, that several assessed vulnerabilities will have the same GOE score. By incorporating CVSS or similar systems, the ranking can be made more granular. However, further research is needed if GOE is supposed to get combined with CVSS in such a way that the modified version of CVSS is meant to produce a value between $0$ (not critical) and $10$ (highest criticality) just like in the default CVSS, but also considers exploitability through AI-driven methods. A discussion within the community is desirable to determine how much a low overall GOE score increases the underlying CVSS score.\par 
The current version of GOE estimates the AI usage based on its effort, the effort is described as how difficult it is to use AI for the respective kill chain step.

The example scores of the last two sections demonstrate that using the GOE is straightforward. The most complex aspect of using it is the research required to assess the questions objectively. Its results further enable security researchers and analysts to strengthen their arguments about why a CVE is easily exploitable via OAI and help them stay updated on the OAI risks associated with a given CVE.
Furthermore, it enhances their argumentation for prioritizing mitigating the risk of a vulnerability by means of quantification. By assessing the GOE analysts can get a rough estimation on how many tools, datasets or APIs for the exploitation of a CVE through OAI exist. Therefore, the GOE also quantifies the AI threat. In \cite{Malatji_2024} mentioned that the quantification of the AI threat is missing in current research. However, it needs to be validated, if the GOEs approach of quantification resembles the reality. Field research needs to be done. To validate the GOEs quantification approach, real life vulnerability exploits through OAI have to be verified. Then, it needs to be validated if a higher amount of tools, datasets or APIs lead to a higher amount of CVE exploitation through OAI.

Another implication of our GOE is that it may be necessary to adapt the vulnerability assessment process. Future security analysts will need a broad understanding of AI and the data required to train AI models. However, current teams may lack it. Therefore, it should be investigated whether it is reasonable to augment current vulnerability assessment teams with AI experts who can provide this comprehensive knowledge and help in assessing the GOE.

To extend and conclude, recent research has increasingly focused on threat modeling for the security of deployed AI assets. This work aims to extend existing threat modeling systems by providing an addition that can be used in vulnerability assessments, specifically addressing the threat of OAI against potentially all assets within an organization —- an emerging research area still in its early stages. We hope that this work encourages further research in the field of threat modeling for OAI used to exploit vulnerabilities across all IT system assets.

\printbibliography
\end{document}